\documentstyle[psfig,graphicx]{mn}

\newif\ifAMStwofonts

\def\simlt{\lower.5ex\hbox{$\; \buildrel < \over \sim \;$}}
\def\simgt{\lower.5ex\hbox{$\; \buildrel > \over \sim \;$}}

\ifoldfss
  \ifCUPmtlplainloaded \else
    \NewTextAlphabet{textbfit} {cmbxti10} {}
    \NewTextAlphabet{textbfss} {cmssbx10} {}
    \NewMathAlphabet{mathbfit} {cmbxti10} {} 
    \NewMathAlphabet{mathbfss} {cmssbx10} {} 
  \fi
  \ifAMStwofonts
    \ifCUPmtlplainloaded \else
      \NewSymbolFont{upmath} {eurm10}
      \NewSymbolFont{AMSa} {msam10}
      \NewMathSymbol{\upi}     {0}{upmath}{19}
      \NewMathSymbol{\umu}     {0}{upmath}{16}
      \NewMathSymbol{\upartial}{0}{upmath}{40}
      \NewMathSymbol{\leqslant}{3}{AMSa}{36}
      \NewMathSymbol{\geqslant}{3}{AMSa}{3E}

    \fi
  \fi
\fi 

\ifnfssone
  \newmathalphabet{\mathit}
  \addtoversion{normal}{\mathit}{cmr}{m}{it}
  \addtoversion{bold}{\mathit}{cmr}{bx}{it}
  \newmathalphabet{\mathbfit} 
  \addtoversion{normal}{\mathbfit}{cmr}{bx}{it}
  \addtoversion{bold}{\mathbfit}{cmr}{bx}{it}
  \newmathalphabet{\mathbfss} 
  \addtoversion{normal}{\mathbfss}{cmss}{bx}{n}
  \addtoversion{bold}{\mathbfss}{cmss}{bx}{n}
  \ifAMStwofonts
    \ifCUPmtlplainloaded \else
      \UseAMStwoboldmath
      \makeatletter
      \new@mathgroup\upmath@group
      \define@mathgroup\mv@normal\upmath@group{eur}{m}{n}
      \define@mathgroup\mv@bold\upmath@group{eur}{b}{n}
      \edef\UPM{\hexnumber\upmath@group}
      \new@mathgroup\amsa@group
      \define@mathgroup\mv@normal\amsa@group{msa}{m}{n}
      \define@mathgroup\mv@bold\amsa@group{msa}{m}{n}
      \edef\AMSa{\hexnumber\amsa@group}
      \makeatother
      \mathchardef\upi="0\UPM19
      \mathchardef\umu="0\UPM16
      \mathchardef\upartial="0\UPM40
      \mathchardef\leqslant="3\AMSa36
      \mathchardef\geqslant="3\AMSa3E
    \fi
  \fi
\fi 

\ifnfsstwo
  \DeclareMathAlphabet{\mathbfit}{OT1}{cmr}{bx}{it}
  \SetMathAlphabet\mathbfit{bold}{OT1}{cmr}{bx}{it}
  \DeclareMathAlphabet{\mathbfss}{OT1}{cmss}{bx}{n}
  \SetMathAlphabet\mathbfss{bold}{OT1}{cmss}{bx}{n}
  \ifAMStwofonts
    \ifCUPmtlplainloaded \else
      \DeclareSymbolFont{UPM}{U}{eur}{m}{n}
      \SetSymbolFont{UPM}{bold}{U}{eur}{b}{n}
      \DeclareSymbolFont{AMSa}{U}{msa}{m}{n}
      \DeclareMathSymbol{\upi}{0}{UPM}{"19}
      \DeclareMathSymbol{\umu}{0}{UPM}{"16}
      \DeclareMathSymbol{\upartial}{0}{UPM}{"40}
      \DeclareMathSymbol{\leqslant}{3}{AMSa}{"36}
      \DeclareMathSymbol{\geqslant}{3}{AMSa}{"3E}
    \fi
  \fi
\fi 

\ifCUPmtlplainloaded \else
  \ifAMStwofonts \else 
    \def\upi{\pi}
    \def\umu{\mu}
    \def\upartial{\partial}
  \fi
\fi

\pagerange {1--9}

\title[Cyclical period changes in V2051 Oph and V4140 Sgr]
	{Cyclical period changes in the dwarf novae V2051 Oph and V4140 Sgr
\thanks{email: bap@astro.ufsc.br~(RB); bernardo@astro.ufsc.br (BWB);
				bond@stsci.edu (HEB); chico@das.inpe.br (FJJ); 
				steiner@astro.iag.usp.br (JES); adgrauer@ualr.edu (ADG)} }
\author[R. Baptista et~al.]
       {R. Baptista $^1$, B. W. Borges $^1$, H. E. Bond $^2$, \cr
        F. Jablonski $^3$, J. E. Steiner $^4$ and A. D. Grauer $^5$ \\
       $^1$ Departamento de F\'{\i}sica, UFSC, Campus Trindade, 88040-900,
       Florian\'opolis, Brazil \\
       $^2$ Space Telescope Science Institute, 3700 San Martin Drive,
       Baltimore, MD 21218, USA \\
       $^3$ Divis\~ao de Astrof\'{\i}sica, Instituto Nacional de Pesquisas
       Espaciais, S.\ J.\ dos Campos, Brazil \\
       $^4$ Departamento de Astronomia - IAG, Universidade de S\~ao Paulo,
       S\~ao Paulo, Brazil \\
       $^5$ Dept.\ of Physics \& Astronomy, University of Arkansas, Little
       Rock, AR 72204, USA }

\date{Accepted 2003 July 15. Received 2003 July 7; in original form 2003 April 10}
\pubyear{2003}

\begin{document}

\maketitle

\begin{abstract}
We report the identification of cyclical changes in the orbital period of
the eclipsing dwarf novae V2051~Ophiuchi and V4140~Sagitarii. 
We used sets of white dwarf mid-eclipse timings to construct
observed-minus-calculated diagrams covering, respectively, 25 and 16 years
of observations. The V2051~Oph data present cyclical variations that 
can be fitted by a linear plus sinusoidal function with period 
$22\pm 2$~yr and amplitude $17\pm 3$~s. The statistical significance of 
this period by an F-test is larger than 99.9 per cent.
The V4140~Sgr data present cyclical variations of similar amplitude and
period $6.9\pm 0.3$~yr which are statistically significant at the 99.7
per cent level.
We derive upper limits for secular period changes of 
$| \hbox{\.{P}} | < 3 \times 10^{-12}$ and 
$| \hbox{\.{P}} | < 1.8 \times 10^{-11}$, respectively
for V2051~Oph and V4140~Sgr.

We combined our results with those in the literature to construct a
diagram of the amplitude versus period of the modulation for a sample
of 11 eclipsing cataclysmic variables (CVs). 
If the cyclical period changes are the consequence of a solar-type 
magnetic activity cycle in the secondary star, then magnetic activity 
is a widespread phenomenon in CVs, being equally common among long- and
short-period systems. This gives independent evidence that the magnetic
field (and activity) of the secondary stars of CVs do not disappear
when they become fully convective. We also find that
the fractional cycle period changes of the short-period CVs are 
systematically smaller than those of the long-period CVs.

\end{abstract}


\section{Introduction}

Cataclysmic variables (CVs) are mass-transferring binary systems 
containing a white dwarf accretor (the primary) and a late-type donor
star (the secondary). If the white dwarf is not strongly magnetized 
($B \simlt 10^6$~Gauss) the transferred matter forms an accretion disc
with a bright spot at the position where the stream of matter hits 
the outer edge of the disc.

In most CVs the donor star has lower mass than the accreting star. 
Since conservative mass transfer in such situations
would lead to an increase in the orbital separation (and therefore
the cessation of mass transfer via Roche lobe overflow), the existence 
of CVs as mass-transfer binaries implies that they must continuously 
lose angular momentum in order to sustain the mass transfer process. 
As a consequence, the binary should evolve slowly towards shorter 
orbital periods (on time scales of $10^8-10^9$~yr). 
Possible mechanisms suggested for driving the continuous angular 
momentum loss are magnetic braking via the secondary star's wind 
(for $P_{orb}>3$ hr) and gravitational radiation (for $P_{orb}< 3$ hr) 
(King 1988).
At very short periods, when the secondary star becomes fully degenerate 
($M_2\simlt 0.08\;M_\odot$), mass loss leads to an expansion of this 
star and reverses the secular trend, resulting thereafter in an increasing
orbital period. However, the predicted mass transfer rate after this 
period minimum is low (\.{M}$_2 \simeq 10^{-12} \; M_\odot\; yr^{-1}$) 
and few CVs are expected to be observed in such evolutionary stage 
(Warner 1995).

CVs show a bi-modal distribution of orbital periods, with systems 
clustering at periods of $1.5-2$~hr or $3-5$~hr, and a remarkable dearth
of systems with periods between $2-3$~hr. This {\em period gap} is
explained in terms of the secular evolution of the binary by the
interrupted braking model (e.g., Hameury, King \& Lasota 1991): 
At periods above the gap, relatively rapid loss of angular momentum 
(via magnetic wind) drives the donor star out of thermal equilibrium 
and makes it oversized for its mass compared with its main sequence radius. 
When the period reaches 3~hr, the angular momentum losses are sharply
reduced, possibly in consequence of a reduced number of open field
lines as the star becomes fully convective. The star then shrinks back 
to its main sequence radius and detaches from the Roche lobe, halting 
mass transfer. Angular momentum losses continue, but on a longer time 
scale set by gravitational radiation, and reduces the binary separation 
until at $P_{orb} \simeq 2$~hr the Roche lobe eventually closes down on 
the star and mass transfer resumes at lower \.{M}.

The secular evolution of the binary can in principle be detected by 
measuring the changes in the orbital period of eclipsing CVs. 
Eclipses  provide a fiducial mark in time and can usually be used to 
determine the orbital period (and its derivative) with high precision.
However, attempts to measure the long-term orbital period decrease in 
CVs have been disappointing: none of the studied stars show a positive 
detection of an orbital period decrease (e.g., Beuermann \& Pakull 1984).
Instead, essencially all of the well observed eclipsing CVs 
\footnote{i.e., those with well-sampled observed-minus-calculated (O$-$C) 
eclipse timings diagram covering more than a decade of observations.}
show cyclical period changes 
(e.g., Bond \& Freeth 1988; Warner 1988; Robinson, Shetrone \& Africano 
1991; Baptista, Jablonski \& Steiner 1992; Echevarria \& Alvares 1993; 
Wolf et~al. 1993; Baptista et~al. 1995; Baptista, Catal\'an \& Costa 2000;
Baptista et~al. 2002). 

The most promising explanation of this effect seems to be the existence 
of a solar-type (quasi- and/or multi-periodic) magnetic activity cycle 
in the secondary star modulating the radius of its Roche lobe and, via
gravitational coupling, the orbital period on time scales of the order 
of a decade (Applegate 1992; Richman, Applegate \& Patterson 1994).
The relatively large amplitude of these cyclical period changes
probably contributes to mask the low amplitude, secular period decrease.

In this paper we report the results of an investigation of orbital 
period changes in the short-period CVs V2051~Oph and V4140~Sgr.
The revised (O$-$C) diagrams show cyclical period changes similar to
those observed in many other well studied eclipsing CVs. The observations
and data analysis are presented in section \ref{observa} and the
results are discussed and summarized in section \ref{discuss}.

\section{Observations and data analysis} \label{observa}

\subsection{V2051 Oph}  \label{oph}

Our observations of V2051\,Oph date back to the night in 1977 when the
eclipses were first discovered by Bond (1977) with one of the 0.9-m
telescopes at Kitt Peak National Observatory (KPNO) (Actually, the very
first eclipse was seen visually by Bond when V2051~Oph disappeared
while he was identifying the star at the eyepiece!). Between 1977 and
1979, we timed eclipses of V2051~Oph with the 0.9-m telescopes at KPNO,
Cerro Tololo Interamerican Observatory (CTIO), and the Louisiana State
University (LSU) Observatory. Most of the data comprises observations
collected with the 1.6\,m telescope at Laborat\'{o}rio Nacional de 
Astrof\'\i sica (LNA) in Brazil from 1981 to 2002. We also observed
V2051\,Oph with the Hubble Space Telescope (HST) on 1996 January 29.

All the KPNO, CTIO and LSU runs, as well as the LNA observations prior to 
1998, were made with photoelectric photometers in a time-series acquisition 
mode using bi-alkali or GaAs photomultiplier tubes.
The data include runs in white light (W) and in the Johnson-Cousins
UBVRI system.  The raw data were sky-subtracted.
Correction for extinction and transparency variations at the LSU 
Observatory was done based on simultaneous measurements of a nearby 
comparison star, using a two-star high-speed photometer (see Grauer
\& Bond 1981).  For the other runs, these effects were corrected 
either by frequent observations of a close comparison star or by
using extinction coefficients calculated from observations of 
photometric standard stars of Graham (1982) and Landolt (1983).
The UBVRI observations were reduced to the standard system by means of 
transformation coefficients for each night, derived from observations of 
blue spectrophotometric standards (Stone \& Baldwin 1983) and standard
stars of Landolt (1983) and Graham (1982). We used the relations
of Lamla (1981) to transform UBVRI magnitudes to flux units.

The remaining LNA observations (from 1998 onward) were made in the 
$B$-band with a high-speed CCD photometer.  Data reduction for these 
data sets included bias subtraction, flat-field correction, cosmic rays 
removal, aperture photometry extraction and absolute flux calibration. 
Time-series were constructed by computing the magnitude difference 
between the variable and a reference comparison star.

The HST runs consist of time-resolved spectroscopy of two consecutive 
eclipses of V2051~Oph collected with the Faint Object Spectrograph (FOS)
on 1996 January 29.  The reader is referred to Baptista et~al. (1998) for 
a detailed description of the data set and of the reduction procedures.

For the photoelectric observations, the internal clock of the photometers
were manually synchronized to UTC time from a WWV radio signal every night
to a precision better than 0.5\,s.  The absolute timing accuracy of the
HST/FOS observations is better than 0.25\,s.  The CCD photometer has a
GPS board which sets its internal clock to UTC time to a precision 
better than 10\,ms.

A summary of the observations is given in Table~\ref{oph.dados}.
The third column gives the time resolution of the observations in seconds.
The fourth column lists the eclipse cycle number.
The data comprise 80 eclipse light curves spanning 25~yr of observations.
All data sets were obtained while V2051~Oph was in quiescence.

We assigned cycle zero to our first eclipse. Thus, in our cycle counting 
scheme the first eclipse in the ephemeris of Bauermann \& Pakull (1984) 
is cycle 12778, while that in the ephemeris of Warner \& O'Donoghue
(1987) corresponds to cycle 24690.
A more complete analysis of the optical data will be presented in a 
separate paper. Here we will concentrate on the measurement of the 
mid-eclipse times.
%
\begin{table}
\begin{minipage}{80mm}
  \caption{Journal of the observations of V2051~Oph.} \label{oph.dados}
\begin{tabular}{@{}lcccl@{}}
Date (UT)   && Passband & $\Delta t$ (s) & Cycles \\
1977 Apr 12 &&    W   & 27.7 & 0 \\
1977 Apr 13 &&    W   & 12.5 & 15, 16 \\
1977 Apr 15 &&    W   &  6.4 & 47, 48 \\
1977 Jun 22 &&    W   & 12.1 & 1\,134 \\
1977 Jun 24 &&    W   & 12.5 & 1\,164, 1\,165 \\
1977 Sep 11 &&    W   & 10.7 & 2\,430 \\
1977 Sep 12 &&    W   & 10.3 & 2\,446 \\
1978 Apr 16 &&    W   & 10.7 & 5\,911 \\
1979 Jun 19 &&    W   &  5.8 & 12\,778, 12\,779 \\
1981 Mar 11 &&    W   &  10  & 22\,887, 22\,888 \\
1981 Jun 10 &&    W   &  10  & 24\,342 \\
1982 May 16 &&    W   &  10  & 29\,789 \\
1988 May 11 &&    W   & 25.2 & 64\,820 \\
1988 Jul 14 &&    W   & 19.3 & 65\,844, 65\,845 \\
1989 Apr 11 && WUBVRI &   3  & 70\,188-70\,190 \\
1989 Jun 06 && WUBVRI &   3  & 71\,082-71\,084 \\
1989 Jun 07 && WUBVRI &   3  & 71\,089 \\
            && WUBVRI &   3  & 71\,099, 71\,100 \\
1989 Jun 08 && WUBVRI &   3  & 71\,114, 71\,115 \\
1989 Aug 04 && WUBVRI &  10  & 72\,028 \\
1990 Mar 26 && WUBVRI &   3  & 75\,781 \\
1990 Aug 13 && WUBVRI &   3  & 78\,034, 78\,036 \\
1990 Aug 14 && WUBVRI &   3  & 78\,050 \\
1990 Aug 15 && WUBVRI &   3  & 78\,066 \\
1991 Mar 15 &&    W   &   5  & 81\,451 \\
1991 Mar 17 &&    W   &  10  & 81\,482 \\
1991 Mar 18 &&    W   &   5  & 81\,498, 81\,499 \\
1991 Sep 04 && WUBVRI &   3  & 84\,233 \\
1991 Sep 05 && WUBVRI &   3  & 84\,249 \\
1992 Jul 29 &&    V   &  30  & 89\,503-89\,505 \\
1996 Jan 29 && G400H  &  3.4 & 109\,988 \\
            && G160L  &  3.4 & 109\,989 \\
1996 Jun 15 &&    B   &   5  & 112\,201 \\
1996 Jun 16 &&    W   &   5  & 112\,217, 112\,218 \\
1996 Jun 17 &&    W   &   5  & 112\,233 \\
1998 Jul 25 &&    B   &   5  & 124\,535-124\,537 \\
1998 Jul 26 &&    B   &   5  & 124\,551-124\,555 \\
1999 Jul 12 &&    B   &   5  & 130\,174, 130\,177 \\
1999 Jul 15 &&    B   &   5  & 130\,222-130\,223 \\
2000 Jul 28 &&    B   &   5  & 136\,293-136\,295 \\
2001 Jun 25 &&    B   &   5  & 141\,611, 141\,612 \\
            &&    B   &  10  & 141\,613, 141\,615 \\
2001 Jun 27 &&    B   &  10  & 141\,643-141\,645 \\
            &&    B   &  10  & 141\,648 \\
2001 Jun 28 &&    B   &  15  & 141\,660, 141\,661 \\
2002 Aug 04 &&    B   &  10  & 148\,099, 148\,100 \\
2002 Aug 07 &&    B   &  20  & 148\,147 \\
\end{tabular}
\end{minipage}
\end{table}

Mid-eclipse times were measured from the mid-ingress and mid-egress
times of the white dwarf eclipse using the derivative technique described
by Wood, Irwin \& Pringle (1985). 
For a given observational season, all light curves were phase-folded 
according to a test ephemeris and sorted in phase to produce a combined 
light curve with increased phase resolution. The combined light curve is 
smoothed with a median filter and its numerical derivative is calculated. 
A median-filtered version of the derivative curve is then analyzed by an
algorithm which identifies the points of extrema (the mid-ingress\,/\,egress
phases of the white dwarf). The mid-eclipse phase, $\phi_0$, is the mean 
of the two measured phases. 
In all cases the difference between the measured mid-egress and mid-ingress
phases is consistent with the expected width of the white dwarf eclipse,
$\Delta\phi= 0.0662 \pm 0.0002$ cycle (Baptista et~al. 1998).
Finally, we adopt a cycle number representative of the ensemble of light
curves and use the same test ephemeris to compute the corresponding observed
mid-eclipse time (HJD) for this cycle including the measured value of $\phi_0$. 
This yields a single,  but robust mid-eclipse timing estimate from a 
sample of eclipse light curves. 
For V2051~Oph these measurements have a typical accuracy of about 5~s.

\begin{figure}	
\includegraphics[bb=1.5cm 3.5cm 18cm 24.5cm,scale=0.47]{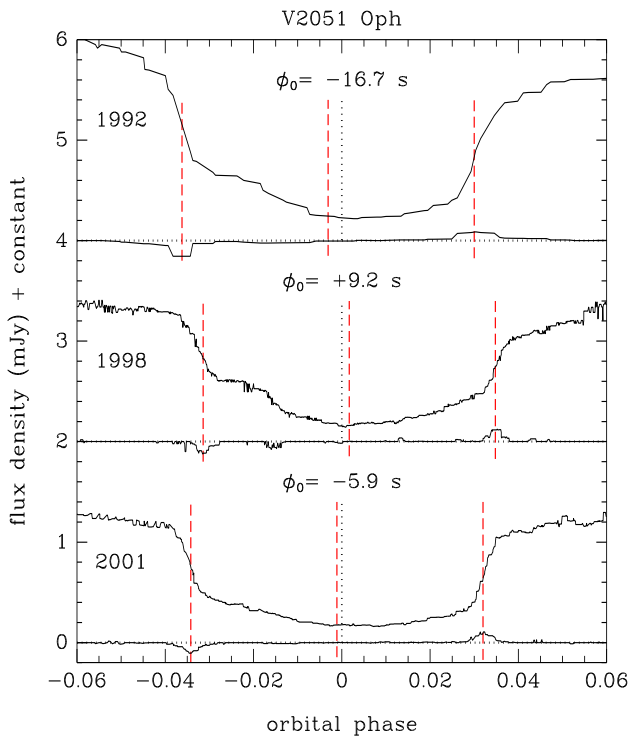}
\caption{ Measuring eclipse timings with the derivative technique.
   The panels show median-filtered average light curves and
   corresponding median-filtered derivative curves of V2051 Oph 
   for 1992 (top), 1998 (middle) and 2001 (bottom). The 1992 and 1998
   curves were vertically displaced by, respectively, 2 and 4 mJy
   for visualization purposes. Horizontal dotted lines indicate the 
   true zero level in each case.
   Vertical dashed lines mark the phases of minimum/maximum derivative 
   (mid- ingress/egress of the white dwarf) and the mid-eclipse	phase,
   $\phi_0$. A vertical dotted line depicts phase zero. The light curves
   were phase-folded according to the best-fit linear ephemeris of 
   Table~\ref{oph.efem}. The time difference between mid-eclipse and 
   phase zero is indicated in each panel. }
  \label{fig1}
\end{figure}

Figure~\ref{fig1} shows the measurements of mid-eclipse timings with
this procedure on average light curves of V2051~Oph at three different
occasions. The sharp breaks in the slope of the light curves correspond to
the ingress/egress features of the white dwarf and provide a precise 
determination of the mid-eclipse phase $\phi_0$.
%
%

The mid-eclipse timings are independent of passband.
The measured timings from the multicolour runs are the same at all
wavelengths within the measurement errors.

For V2051~Oph the difference between universal time (UT) and terrestrial 
dynamical time (TDT) changed by 14~s over the 26~yr time span of our 
observations.  The amplitude of the difference between the baricentric 
and the heliocentric corrections is about 7~s.
The mid-eclipse timings have been calculated on the solar system baricentre
dynamical time (BJDD), according to the code by Stumpff (1980).
The terrestrial dynamical (TDT) and ephemeris (ET) time scales were 
assumed to form a contiguous scale for our purposes.
The new eclipse timings are listed in Table~\ref{oph.timings}.
The corresponding uncertainties in the last digit are indicated in 
parenthesis.
%
%
\begin{table}
\begin{minipage}{80mm}
  \caption{Mid-eclipse timings of V2051~Oph.} \label{oph.timings}
\begin{tabular}{@{}rccl@{}}
\multicolumn{1}{c}{Cycle} & HJD & BJDD & ~(O$-$C) 
\footnote{Observed minus calculated times with respect to the linear 
ephemeris of Table~\ref{oph.efem}.} \\ [-0.5ex]
     & (2,400,000+) & (2,400,000+) & ~(cycles) \\ [1ex]
      24 & 43247.47528 & 43247.47586 (5) & $+0.0011$ \\ 
  1\,154 & 43318.01879 & 43318.01937 (5) & $+0.0015$ \\ 
  2\,038 & 43373.20507 & 43373.20565 (6) & $+0.0023$ \\ 
  5\,911 & 43614.98815 & 43614.98874 (6) & $+0.0018$ \\ 
 12\,778 & 44043.68031 & 44043.68089 (7) & $+0.0022$ \\ 
 22\,888 & 44674.82576 & 44674.82633 (6) & $-0.0020$ \\ 
 24\,342 & 44765.59587 & 44765.59644 (6) & $-0.0021$ \\ 
 29\,789 & 45105.64044 & 45105.64100 (9) & $-0.0022$ \\ 
 64\,820 & 47292.55070 & 47292.55135 (6) & $-0.0044$ \\ 
 65\,845 & 47356.53932 & 47356.53997 (5) & $-0.0034$ \\ 
 70\,999 & 47678.29257 & 47678.29323 (6) & $-0.0026$ \\ 
 77\,594 & 48090.00424 & 48090.00491 (6) & $-0.0039$ \\ 
 81\,482 & 48332.72384 & 48332.72451 (7) & $-0.0029$ \\ 
 84\,241 & 48504.96234 & 48504.96301 (6) & $-0.0025$ \\ 
 89\,504 & 48833.52016 & 48833.52082 (5) & $-0.0030$ \\ 
109\,989 & 50112.35515 & 50112.35581 (4) & $+0.0003$ \\ 
112\,217 & 50251.44442 & 50251.44508 (2) & $+0.0002$ \\ 
124\,545 & 51021.05518 & 51021.05589 (2) & $+0.0018$ \\ 
130\,199 & 51374.02226 & 51374.02299 (2) & $+0.0012$ \\ 
136\,294 & 51754.52000 & 51754.52074 (4) & $+0.0000$ \\ 
141\,636 & 52088.00958 & 52088.01033 (2) & $-0.0010$ \\ 
148\,100 & 52491.54297 & 52491.54371 (6) & $-0.0061$ \\ [-4ex]
\end{tabular}
\end{minipage}
\end{table}

The data points were weighted by the inverse of the squares of the
uncertainties in the mid-eclipse times. 
Table~\ref{oph.efem} presents the parameters of the best-fit linear, 
quadratic and linear plus sinusoidal ephemerides with their 1-$\sigma$ 
formal errors quoted.
We also list the root-mean-squares of the residuals, $\sigma$, and the 
$\chi^{2}_{\nu_{2}}$ value for each case, where $\nu_{2}$ is the 
number of degrees of freedom.
%
%
\begin{table}
\centering
 \begin{minipage}{80mm}
 \caption{Ephemerides of V2051~Oph.}  \label{oph.efem}
\begin{tabular}{@{}ll@{}}

\multicolumn{2}{l}{\bf Linear ephemeris:} \\
\multicolumn{2}{l}{BJDD = T$_{0}$ + P$_{0}\cdot E$} \\ [1ex]
T$_{0} = 2\,443\,245.977\,52\,(\pm 3)$ d &
P$_{0} = 0.062\,427\,8634\,(\pm 3)$ d \\
$\chi^{2}_{\nu_{2}}= 9.5, \;\;\; \nu_{2}$ = 20 &
$\sigma_{1}= 2.69 \times 10^{-3}$ cycles \\ [2ex]

\multicolumn{2}{l}{\bf Quadratic ephemeris:} \\ 
\multicolumn{2}{l}{BJDD = T$_{0}$ + P$_{0}\cdot E$ + $c\cdot E^{2}$} \\ [1ex]
T$_{0} = 2\,443\,245.977\,59\,(\pm 3)$ d & 
P$_{0} = 0.062\,427\,859\,(\pm 1)$ d \\
c $ = (+3.2 \pm 0.8) \times 10^{-14}$ d &
$\sigma_{2} = 2.29 \times 10^{-3}$ cycles \\ 
$\chi^{2}_{\nu_{2}}= 8.5 , \;\;\; \nu_{2}$ = 19 \\ [2ex]

\multicolumn{2}{l}{\bf Sinusoidal ephemeris:} \\
\multicolumn{2}{l}{BJDD = T$_{0}$ + P$_{0}\cdot E$ + A$\cdot 
\cos\,[2\pi (E-{\rm B})/{\rm C}]$} \\ [1ex]
T$_{0}= 2\,443\,245.977\,45\,(\pm 4)$ d & 
B $= (120 \pm 6) \times 10^{3}$ cycles \\
P$_{0}= 0.062\,427\,8629\,(\pm 4)$ d  & 
C $= (127 \pm 8) \times 10^{3}$ cycles \\
A $= (20 \pm 4) \times 10^{-5}$ d & 
$\sigma_{\rm S}= 1.22 \times 10^{-3}$ cycles \\
$\chi^{2}_{\nu_{2}}= 2.5, \;\;\; \nu_{2}= 17$ \\
\end{tabular}
\end{minipage}
\end{table}
%

\begin{figure}	
\includegraphics[bb=1.1cm 1.5cm 18cm 27cm,scale=0.45]{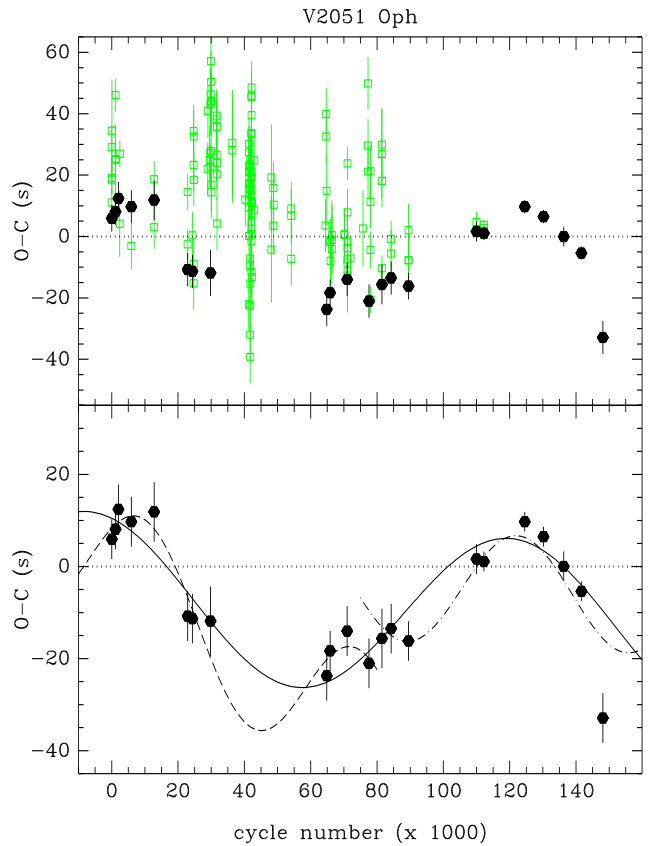}
\caption{ The (O$-$C) diagram of V2051~Oph with respect to the linear 
  ephemeris of Table~\ref{oph.efem}. 
  The eclipse timings from the literature and individual timings 
  measured from our light curves using the bisected chord method are
  shown as open squares, while the new mid-eclipse timings are 
  indicated as solid circles.
  The solid line in the lower panel shows the best-fit linear plus 
  sinusoidal ephemeris of Table~\ref{oph.efem}.  The dashed and 
  dot-dashed lines in the lower panel show the best-fit 11~yr cycle
  period sinusoidal ephemeris, respectively, for the data in the first 
  and the second halves of the time interval. }
  \label{fig2}
\end{figure}

Fig.~\ref{fig2} presents the (O$-$C) diagram with respect to the linear 
ephemeris in Table~\ref{oph.efem}. The new timings are indicated as 
solid circles and show a clear modulation. 
Open squares show the individual eclipse timings taken from the 
literature (Warner \& Cropper 1983; Cook \& Brunt 1983; Bauermann \& 
Pakull 1984; Watts et al. 1986; Watts \& Watson 1986; Warner \& 
O'Donoghue 1987; Hollander, Kraakman \& van Paradijs 1993; Echevarria 
\& Alvares 1993; Baptista et~al. 1998) and individual eclipse timings 
measured from our light curves using the bisected chord method.
All individual timings were corrected to BJDD.
The individual timings show a large scatter, of about 40~s, and are 
generally later than the contemporary mid-eclipse timings in 
Table~\ref{oph.timings}. We will return to this point in 
section~\ref{discuss}.
%
%

The significance of adding additional terms to the linear ephemeris 
was estimated by using the F-test, following the prescription of 
Pringle (1975). The quadratic ephemeris has a statistical significance 
of 98.6 per cent with an $F(3,19) = 7.2$.
On the other hand, the statistical significance of the linear plus 
sinusoidal ephemeris with respect to the linear fit is larger than 99.99 
per cent, with an $F(3,17) = 65.6$. 
The best-fit period of the modulation is $22 \pm 2$~yr.
The best-fit linear plus sinusoidal ephemeris is shown as a solid line in 
the lower panel of Fig.~\ref{fig2}.

However, the eclipse timings show systematic and significant deviations 
from the best-fit linear plus sinusoidal ephemeris.  The fact that
$\chi^{2}_{\nu_{2}}>1$ emphasizes that the linear plus sinusoidal 
ephemeris is not a complete description of the data, perhaps signalling 
that the period variation is not sinusoidal or not strictly periodic.

We explored these possibilities by making separate fits to the first 
half of the data set ($E<75\times 10^3$ cycle) and to the second half
of the data set ($E>75 \times 10^3$ cycle).
The first half of the data set can be well described by ephemerides 
with cycle periods of 11 and 5.5~yr.
The best-fit cycle period for the second half of the data set is the 
same as that obtained for the whole data set under the uncertainties. 
This part of the data set can also be reasonably well described by
ephemerides of cycle periods of 11 and 7~yr.
The 7~yr period is consistent with the 6.87~yr cycle period previously
found by Echevarria \& Alvares (1993).
We note that 11, 7 and 5.5~yr periods are, respectively, the first, 
second and third harmonics of the 22~yr main period modulation.
The 11~yr period best-fit ephemerides for the first and second halves
of the data set are shown in the lower panel of Fig.~\ref{fig1}, 
respectively, as the dashed and dot-dashed curves.
These results indicate that the period changes in V2051~Oph are not 
sinusoidal or not strictly periodic.

\subsection{V4140~Sgr}  \label{sgr}

V4140~Sgr was observed with the 1.6\,m telescope at LNA from 1985 to 2000.
The reader is referred to Baptista et~al. (1989; 1992) for a detailed 
description of the data sets and of the reduction procedures for the
observations prior to 1992. The new observations (1992-2000) were 
performed with the same CCD photometer as those of V2051~Oph except 
for the run of 1996 June, which was collected with the photoelectric
photometer. The data reduction in each case is identical to that 
described in section~\ref{oph}.

The journal of the new observations is shown in Table~\ref{sgr.dados}.
The notation is the same as in Table~\ref{oph.dados}.
The data comprise 57 eclipse light curves spanning 16~yr of observations.
Only runs while V4140~Sgr was in quiescence were included in the 
analysis. The observations of the eclipse cycles 18782-18783, 
18798-18801, and 30228-30229 (Baptista et~al. 1992) were excluded 
because V4140~Sgr was in outburst at those epochs. 
%
\begin{table}
\begin{minipage}{80mm}
  \caption{The new observations of V4140~Sgr.} \label{sgr.dados}
\begin{tabular}{@{}lccl@{}}
Date (UT)   &Passband& $\Delta t$ (s) & Cycle       \\
1992 Jul 29 &  V   &     15        & 41\,867--41870   \\
1992 Jul 30 &  V   &     15        & 41\,882--41887   \\
1996 Jun 16 &  W   &     15        &     64\,955      \\
1998 Jul 26 &  B   &     20        & 77\,488, 77\,489 \\
1999 Jul 12 &  B   &     20        & 83\,200, 83\,202 \\
1999 Jul 14 &  B   &     20        &     83\,236      \\
2000 Jul 29 &  B   &     20        &     89\,436      \\
2000 Jul 30 &  B   &     20        & 89\,449, 89\,451 \\
\end{tabular}
\end{minipage}
\end{table}

Mid-eclipse times of the white dwarf were measured for the ensemble
of light curves of each season with the technique described in
section~\ref{oph}. The accuracy of these measurements for V4140~Sgr
is better than 12~s. The lower accuracy is mainly the consequence of
V4140~Sgr being 3 magnitudes fainter than V2051~Oph.

For V4140~Sgr the difference between the UT and TDT scales changed by
6~s over the 16~yr span of our data set, and the amplitude of the 
difference between the baricentric and the heliocentric corrections 
is about 4~s.
The mid-eclipse timings were calculated in BJDD according to the 
code by Stumpff (1980) and are listed in Table~\ref{sgr.timings}.
The corresponding uncertainties in the last digit are indicated in 
parenthesis.
%
%
\begin{table}
\begin{minipage}{80mm}
  \caption{Mid-eclipse timings of V4140 Sgr.} \label{sgr.timings}
\begin{tabular}{@{}rcll@{}}
\multicolumn{1}{c}{Cycle} & HJD & \multicolumn{1}{c}{BJDD} & ~(O$-$C) 
\footnote{Observed minus calculated times with respect to the 
linear ephemeris of Table~\ref{sgr.efem}.} \\ [-0.5ex]
      & (2,400,000+)& ~~~(2,400,000+)   & ~(cycles) \\ [1ex]
     12 & 46262.40830 & 46262.40889 (7)  & $+0.0046$ \\
 5\,179 & 46579.81529 & 46579.81589 (9)  & $+0.0023$ \\
13\,080 & 47065.17109 & 47065.17172 (10) & $+0.0013$ \\
17\,826 & 47356.71619 & 47356.71683 (7)  & $-0.0008$ \\
23\,162 & 47684.50477 & 47684.50543 (6)  & $-0.0036$ \\
30\,065 & 48108.55396 & 48108.55463 (11) & $-0.0013$ \\
35\,580 & 48447.33864 & 48447.33931 (13) & $-0.0013$ \\
41\,878 & 48834.22291 & 48834.22358 (13) & $+0.0014$ \\
64\,955 & 50251.83520 & 50251.83585 (13) & $-0.0053$ \\
77\,488 & 51021.73363 & 51021.73431 (6)  & $-0.0004$ \\
83\,213 & 51373.41854 & 51373.41923 (6)  & $-0.0000$ \\
89\,445 & 51756.24831 & 51756.24903 (4)  & $+0.0006$ \\ [-4ex]
\end{tabular}
\end{minipage}
\end{table}

We fitted linear, quadratic and linear plus sinusoidal ephemerides
to the eclipse timings in Table~\ref{sgr.timings}.
The data points were weighted by the inverse of the squares of the
uncertainties in the mid-eclipse times. 
Table~\ref{sgr.efem} shows the parameters of the best-fit ephemerides 
with their 1-$\sigma$ formal errors quoted.  The values of $\sigma$ 
and $\chi^{2}_{\nu_{2}}$ for each case are also listed.
%
%
\begin{table}
\centering
 \begin{minipage}{80mm}
 \caption{Ephemerides of V4140~Sgr.}  \label{sgr.efem}
\begin{tabular}{@{}ll@{}}

\multicolumn{2}{l}{\bf Linear ephemeris:} \\
\multicolumn{2}{l}{BJDD = T$_{0}$ + P$_{0}\cdot E$} \\ [1ex]
T$_{0} = 2\,446\,261.671\,45\,(\pm 6)$ d &
P$_{0} = 0.061\,429\,6779\,(\pm 9)$ d \\
$\chi^{2}_{\nu_{2}}= 4.2, \;\;\; \nu_{2}$ = 10 &
$\sigma_{1}= 2.62 \times 10^{-3}$ cycles \\ [2ex]

\multicolumn{2}{l}{\bf Quadratic ephemeris:} \\
\multicolumn{2}{l}{BJDD = T$_{0}$ + P$_{0}\cdot E$ + $c\cdot E^{2}$} \\
[1ex]
T$_{0} = 2\,446\,261.671\,68\,(\pm 7)$ d &
P$_{0} = 0.061\,429\,660\,(\pm 5)$ d \\
c $ = (+1.8 \pm 0.5) \times 10^{-13}$ d &
$\sigma_{2} = 1.88 \times 10^{-3}$ cycles \\
$\chi^{2}_{\nu_{2}}= 1.74, \;\;\; \nu_{2}$ = 9 \\ [2ex]

\multicolumn{2}{l}{\bf Sinusoidal ephemeris:} \\
\multicolumn{2}{l}{BJDD = T$_{0}$ + P$_{0}\cdot E$ + A$\cdot
\cos\,[2\pi (E-{\rm B})/{\rm C}]$} \\ [1ex]
T$_{0}= 2\,446\,261.671\,50\,(\pm 4)$ d & B $=(3\pm 2)\times 10^{3}$ cycles \\
P$_{0}= 0.061\,429\,6757\,(\pm 7)$ d & C $=(41\pm 2)\times 10^{3}$ cycles \\
A $= (19 \pm 6) \times 10^{-5}$ d &
$\sigma_{\rm S}= 0.92 \times 10^{-3}$ cycles \\
$\chi^{2}_{\nu_{2}}= 0.71, \;\;\; \nu_{2}= 7$ \\
\end{tabular}
\end{minipage}
\end{table}

\begin{figure}	
\includegraphics[bb=1.1cm 1.4cm 18cm 26.5cm,scale=0.45]{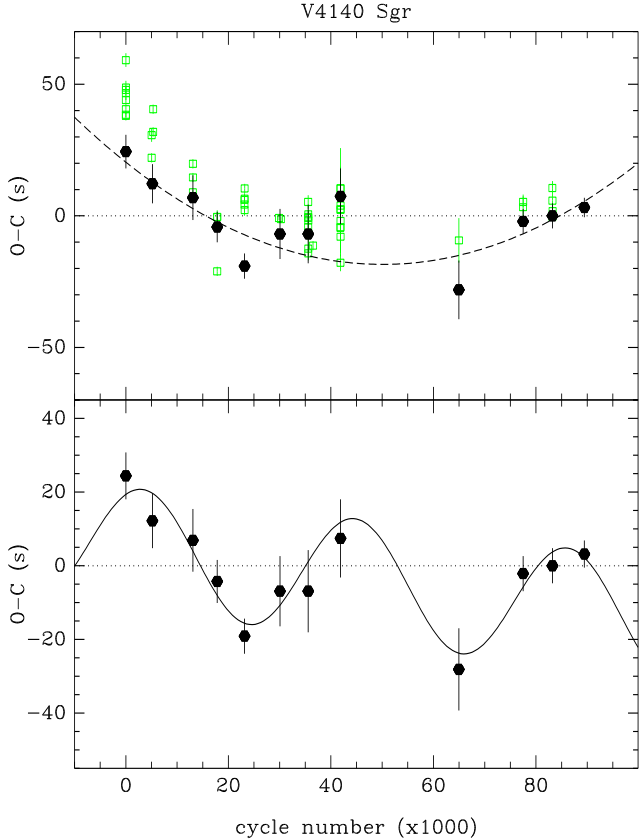}
\caption{ The (O$-$C) diagram of V4140~Sgr with respect to the linear 
  ephemeris of Table~\ref{sgr.efem}. 
  The individual timings from the literature and eclipse timings 
  measured from our individual light curves using the bisected chord 
  method are shown as open squares, while the new timings are indicated 
  as solid circles.
  The dashed line in the upper panel depicts the best-fit quadratic 
  ephemeris while the solid line in the lower panel shows the best-fit 
  linear plus sinusoidal ephemeris of Table~\ref{sgr.efem}. }
  \label{fig3}
\end{figure}

Fig.~\ref{fig3} shows the (O$-$C) diagram with respect to the linear 
ephemeris in Table~\ref{sgr.efem}. The new timings are indicated as 
solid circles. Open squares show the timings from the literature and
individual eclipse timings measured from our new light curves using 
the bisected chord method.  These timings were corrected to BJDD.
Similar to Fig.~\ref{fig2}, the individual timings show a tendency to
be later than the white dwarf mid-eclipse timing of the ensemble.
This effect will be addressed in section~\ref{discuss}.
%

From the $\sigma_1$ and $\sigma_2$ values in Table~\ref{sgr.efem} we
obtain $F(3,9) = 8.5$, which corresponds to a 98.3 per cent confidence
level for the quadratic ephemeris.  This is less than the 99.7 per cent 
confidence level previously found by Baptista et~al. (1992).
The quadratic term is a factor of 2.4 times smaller than that derived 
by those authors.
The best-fit quadratic ephemeris is depicted as a dashed line in the 
upper panel of Fig.~\ref{fig3}.
The linear plus sinusoidal ephemeris is statistically significant at
the 99.97 per cent level for an $F(3,7) = 49.8$. The dispersion of
the data with respect to the fit is reduced by a factor of 2 in
comparison with the quadratic ephemeris.
The best-fit period of the modulation in V4140~Sgr is $6.9 \pm 0.3$~yr.
The best-fit linear plus sinusoidal ephemeris is shown as 
a solid line in the lower panel of Fig.~\ref{fig3}.
A search for a variable cycle period or for harmonics of the main 
cycle period (by performing separate fits to different parts of the 
data set) is not conclusive in this case because of the relatively 
small number of mid-eclipse timings.

\section{Discussion} \label{discuss}

\subsection{On the large dispersion of the V2051~Oph eclipse timings}
\label{dispersion}

V2051~Oph shows large amplitude flickering which distorts its eclipse
shape (e.g., Warner \& Cropper 1983) and affects the measurement 
of eclipse times. The large dispersion in the individual eclipse 
timings in comparison with our results can be partly
attributed to the highly variable shape of the eclipse in V2051~Oph 
and partly to the variety of methods employed by
different authors to measure their eclipse timings. Some of the authors 
measure times of minimum light (e.g., Hollander et~al. 1993), while 
others quote the time corresponding to the centroid of the eclipse
\footnote{usually referred in the literature as the mid-eclipse time.} 
(e.g., Warner \& O'Donoghue 1987). 
None of the individual timings in the literature correspond to the
mid-eclipse of the white dwarf. This is not surprising since the
white dwarf eclipse features are not easy to see in individual light
curves because of flickering.

Eclipses of cataclysmic variables are usually asymmetric because of the
non-negligible contribution of the bright spot 
to the total light. In the case of quiescent dwarf novae -- where the
accretion disc is relatively faint in comparison to the white dwarf 
and the bright spot -- the asymmetry is even more pronounced, leading
to a double-stepped eclipse shape (p.ex., Fig.~\ref{fig1}).

For such asymmetric eclipse shape, both the times of minimum light and
of the centroid of the eclipse will be affected by the relative 
brightness and by the position of the bright spot with respect to the 
line joining the stars.
Times of minimum light tend to be later than the times of the centroid
of the eclipse. Both of them are later than the white dwarf 
mid-eclipse time.
The difference between these three quantities increase with the
relative brightness of the bright spot.

By combining many eclipse light curves we are able of reducing the
influence of flickering in the determination of the eclipse timing.
By applying a unique measurement procedure to all data set and by
adopting the white dwarf mid-eclipse as our timing reference
we are able of minimizing the internal inconsistencies
affecting the individual timings in the literature.

The lesson Fig.~\ref{fig2} tells us is that the
ability to identify small amplitude ($\simlt 40$~s) orbital period
modulations may be considerably limited with (O$-$C) diagrams based 
on individual eclipse timings measured with different methods.

\subsection{Upper limits for secular period changes}

The statistical significance of the quadratic ephemerides of V2051~Oph 
and V4140~Sgr is below the 3-$\sigma$ level and is highly dependent
on the eclipses timings included in the fit. For example, if the
first timing of V4140~Sgr is not included in the fit the significance 
of the resulting quadratic ephemeris drops to 93 per cent.

This behaviour is easy to understand if we assume that the period 
cyclical modulation is real.
In the limit where the observing baseline covers a small number of
cycles of the period modulation, the sign and the magnitude of the 
quadratic term are affected by the number of maxima and minima
comprised by the baseline. In the case of equal number of maxima 
and minima, the quadratic term of a parabolic fit will tend to zero
(or, to the true quadratic trend in the data set).
If there are more maxima (minima) then minima (maxima), the quadratic term
will tend to be positive (negative). For V2051~Oph, the observations cover 
two maxima and one minimum of the modulation and the quadratic term of
the parabolic fit is positive. For V4140~Sgr there are three maxima and
two minima in the baseline, also leading to a positive quadratic term.
Hence, we conclude that there is not yet evidence of secular period
increase/decrease in these two binaries.  

We may use the computed quadratic term to derive a 3-$\sigma$ upper
limit to a secular period change.
For V2051~Oph this gives $| \hbox{\.{P}} | < 3 \times 10^{-12}$, 
corresponding to a timescale for secular period change of 
$\tau_P > 5.6 \times 10^7$~yr.  These numbers are a factor of 
about 5 times more restrictive than those previously derived by 
Echevarria \& Alvares (1993).
For V4140~Sgr we find $| \hbox{\.{P}} | < 1.8 \times 10^{-11}$ and 
$\tau_P > 9.6 \times 10^6$~yr.

An upper limit to the mass transfer rate in the binary can be obtained 
from the timescale $\tau_{P}$. Using a power law to express the response
of the radius of the secondary star to mass loss, R$_{2}(t) \propto 
[{\rm M}_{2}(t)]^\beta$, the mass transfer rate \.{M}$_{2}$ can be 
written in terms of $\tau_P$ as (Molnar 1988; Robinson et~al. 1991),
\begin{equation}
\frac{\hbox{\.{M}}_2}{{\rm M}_2} = \left( \frac{2}{3\beta-1} \right)
\frac{1}{\tau_P} = \frac{1.6}{\tau_P} \;\;\;\; ,
\end{equation}
\noindent
where we assumed $\beta= 0.75$ (Smith \& Dhillon 1998).
Adopting $M_2= 0.15\,M_\odot$ (Baptista et~al. 1998) and 
$M_2= 0.09\,M_\odot$ (Borges \& Baptista 2003, private communication), 
we find \.{M}$_2 < 4.3 \times 10^{-9}\; M_\odot\; yr^{-1}$ and
\.{M}$_2 < 1.5 \times 10^{-8}\; M_\odot\; yr^{-1}$, respectively,
for V2051~Oph and V4140~Sgr.

\subsection{Cyclical period changes}

Our results reveal that the orbital period of V2051~Oph shows conspicuous
cyclical, quasi-periodic changes of amplitude 17~s on a time-scale of
about 22~yr. V4140~Sgr shows a similar period modulation on a time-scale
of 6.9~yr.

Cyclical orbital period changes are seen in many eclipsing CVs (Warner 
1995 and references therein). The cycle periods range from 4~yr in EX~Dra 
(Baptista et~al. 2000) to about 30~yr in UX~UMa (Rubenstein, Patterson 
\& Africano 1991), whereas the amplitudes are in the range $10^1 -10^2$~s. 
Therefore, V2051~Oph and V4140~Sgr fit nicely in the overall picture 
drawn from the observations of orbital period changes in CVs.

If one is to seek for a common explanation for the cyclical period
changes in CVs, then models involving apsidal motion or a third body
in the system shall be discarded as these require that the orbital period 
change be strictly periodic, whereas the observations show that this
is not the case (Richman et~al. 1994 and references therein).
We may also discard explanations involving angular momentum exchange in
the binary, as cyclical exchange of rotational and orbital angular 
momentum (Smak 1972; Biermann \& Hall 1973) requires discs with masses 
far greater than those deduced by direct observations, and 
the time-scales required to allow the spin-orbit coupling of a secondary 
of variable radius are much shorter than the tidal synchronization scales 
for these systems ($\sim 10^{4}$ yr, see Applegate \& Patterson 1987).

The best current explanation for the observed cyclical period modulation 
is that it is the result of a solar-type magnetic activity cycle in the
secondary star (see Applegate 1992 and references therein). 
Richman et~al. (1994) proposed a model in which the Roche lobe radius 
of the secondary star $R_{L2}$ varies in response to changes in the 
distribution of angular momentum inside this star (caused by the magnetic
activity cycle), leading to a change in the orbital separation and, 
therefore, in the orbital period. 
As a consequence of the change in the Roche lobe radius, the mass 
transfer rate \.{M}$_2$ also changes. 
In this model, the orbital period is the shortest when the secondary star 
is the most oblate (i.e., its outer layers rotate faster), and is the 
longest when the outer layers of the secondary star are rotating the 
slowest.

The fractional period change $\Delta P/P$ is related to the amplitude
$\Delta(O-C)$ and to the period $P_{mod}$ of the modulation by 
(Applegate 1992),
\begin{equation}
\frac{\Delta P}{P}= 2\pi\; \frac{\Delta(O-C)}{P_{mod}}= 2\pi\; \frac{A}{C} \; .
\label{eq:pponto}
\end{equation}

Using the values of $A$ and $C$ in Tables~\ref{oph.efem} and \ref{sgr.efem}, 
we find $\Delta P/P = 1.6 \times 10^{-7}$ and $\Delta P/P = 4.7 \times
10^{-7}$, respectively, for V2051~Oph and V4140~Sgr.

The predicted changes in Roche lobe radius and mass transfer rate 
are related to the fractional period change by (Richman et~al. 1994),
\begin{equation}
\frac{\Delta R_{L2}}{R_{L2}}= 39\; \left( \frac{1+q}{q} \right)^{2/3}
\left( \frac{\Delta\Omega}{10^{-3}\Omega} \right)^{-1} \frac{\Delta P}{P} \; ,
\end{equation}
and by,
\begin{equation}
\frac{\Delta \hbox{\.{M}}_2}{\hbox{\.{M}}_2} =
1.22 \times 10^5 \; \left( \frac{1+q}{q} \right)^{2/3}
\left( \frac{\Delta\Omega}{10^{-3}\Omega} \right)^{-1} \frac{\Delta P}{P} \; ,
\end{equation}
where $\Delta\Omega/\Omega$ is the fractional change in the rotation
rate of the outer shell of the secondary star involved in the cyclical 
exchange of angular momentum, $q\; (= M_2/M_1)$ is the binary mass ratio, 
and the minus signs were dropped.

Baptista et~al. (2002) showed that the induced changes in mass transfer 
rate lead to a modulation in the quiescent brightness of a dwarf nova of
$\Delta m \simeq 0.3 \; ( \Delta \hbox{\.{M}}_2 / \hbox{\.{M}}_2 )$.
They inferred a value of $\Delta\Omega/\Omega \simeq 2.7 \times 10^{-3}$ 
for Z~Cha by equating the predicted change in brightness to the observed 
brightness modulation (Ak, Ozkan \& Mattei 2001).
Adopting $q=0.19$ for V2051~Oph (Baptista et~al. 1998) and $q=0.125$ for
V4140~Sgr (Borges \& Baptista 2003, priv. commun.), and assuming 
$\Delta\Omega/\Omega \simeq 3 \times 10^{-3}$, we obtain
$\Delta R_{L2}/R_{L2} \simeq 7 \times 10^{-6}$ and 
$\Delta R_{L2}/R_{L2} \simeq 2.6 \times 10^{-5}$, respectively
for V2051~Oph and V4140~Sgr. 
This is comparable to the value obtained for Z~Cha.
(We found an error in the fractional change of the radius of the
secondary Roche lobe of Z~Cha as quoted by Baptista et~al. (2002).
The correct value is $\Delta R_{L2}/R_{L2}= 2.2 \times 10^{-5}$.)

\subsection{The amplitude versus $P_{mod}$ diagram}

The present work doubles the sample of measured $\Delta P/P$ values
for short-period eclipsing CVs and allows a qualitative comparison of 
cyclical period changes between systems above and below the period gap.

\begin{figure*}	
\includegraphics[bb=1cm 1cm 20.5cm 26.5cm,angle=-90,scale=0.45]{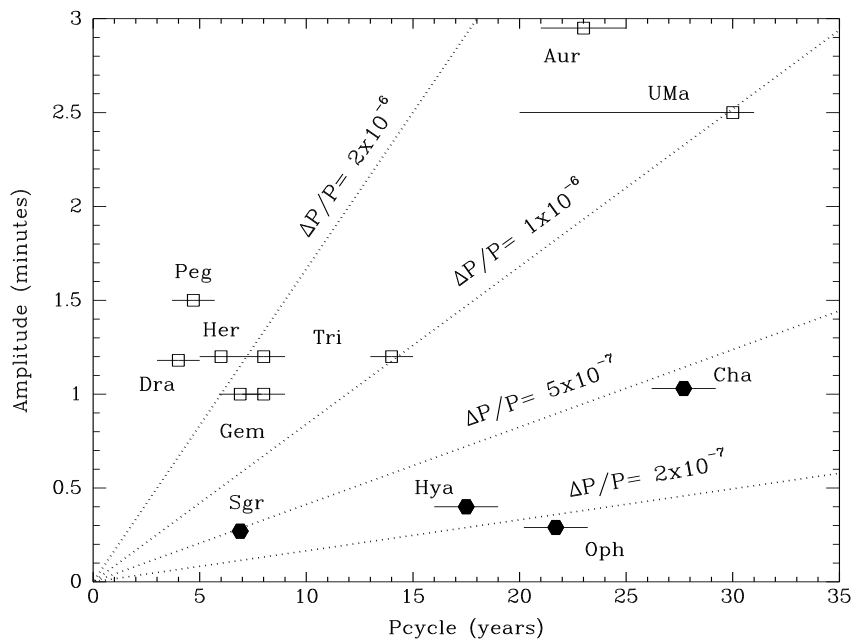}
\caption{ Diagram of the amplitude versus period of the modulation for 
eclipsing CVs. Lines of constant fractional period change $\Delta P/P$
are shown as dotted lines. Long period systems ($P_{orb}>3$~hr) are
shown as open squares, short period system ($P_{orb}>3$~hr) are 
indicated by filled circles. The diagram include measurements of
the following stars: V2051~Oph, V4140~Sgr (this paper), Z~Cha (Baptista
et~al. 2002), EX~Dra (Baptista, Catal\'an \& Costa 2000), and EX~Hya, 
U~Gem, IP~Peg, DQ~Her, RW~Tri, T~Aur and UX~UMa (Warner 1995 and
references therein). }
  \label{fig4}
\end{figure*}

Figure~\ref{fig4} shows a diagram of the amplitude versus period of the
modulation for 11 eclipsing CVs, comprising 7 long-period 
($P_{orb}> 3$~hr) and 4 short-period ($P_{orb}< 3$~hr) systems.
The absence of a few well-known eclipsing CVs in this diagram (e.g.,
HT~Cas and OY~Car) is worth a comment. The lack of evidence of
cyclical period changes in these systems may be caused by selection 
effects, p.ex., (O$-$C) diagrams covering a not long enough baseline
($\simlt 10$~yr) or with a poor sampling of eclipse timings. 
Additionally, small amplitude period modulations may be buried in the
scatter of a sample of individual eclipse timings measured with 
different procedures (section~\ref{dispersion}).
%
%

Under the framework of the magnetic activity cycle explanation,
Fig.~\ref{fig4} allows a few interesting conclusions to be drawn.
First, magnetic activity cycles seems a widespread phenomenon in
the secondary stars of CVs, being equally common among long- and 
short-period systems. Noteworthy, even the fully convective
secondary stars of the short-period CVs show magnetic activity 
cycles. The cycle periods are in the range from 4 to 30~yr, 
independent of binary orbital period.

These results are consistent with those of Ak, Ozkan \& Mattei (2001).
They found cyclical variations in the quiescent magnitude and outburst
interval of a sample of CVs, which they attributed to solar-type
magnetic activity cycles in the secondary stars. They also found
no correlation of the cycle period with the rotation regime of the 
secondary star (i.e., orbital period, for the phase-locked secondary
stars in CVs).

Thus, we collected independent evidence that the magnetic field 
(and activity) of the secondary stars of CVs do not disappear
when they become fully convective. 

Last but not least, the fractional cycle period changes of the 
short-period CVs [$\Delta P/P \simeq (2-5)\times 10^{-7}$] are 
systematically smaller than those of the long-period CVs 
[$\Delta P/P \simeq (1-2) \times 10^{-6}$]. 
Assuming that ($\Delta\Omega / \Omega$) is the same for long- and
short-period systems, this leads to systematically larger fractional 
changes in the Roche lobe of the secondary star for the long period CVs 
(by a factor of $\simeq 2.5$). Alternatively, assuming that the
resulting $\Delta R_{L2}/R_{L2}$ are comparable for long- and 
short-period CVs, the above implies that the ($\Delta\Omega / \Omega$) 
values are systematically larger by a factor of $\simeq 2.5$ for the
long-period systems.
This effect may be a consequence of the different internal structures 
of secondary stars of long- and short-period CVs or it may alternatively
be a manifestation of a lower intensity magnetic field in the 
short-period CVs.

Investigation of period changes in CVs have usually been done by
collecting individual timings in the literature and adding new ones
to construct updated (O$-$C) diagrams. Our analysis of the V2051~Oph
data shows that more elaborated procedures, such as the one
described in this paper, are probably required in order to find
low-amplitude cyclical period changes in short-period CVs.
The extension of the sample of systems in the amplitude versus $P_{mod}$
diagram, particularly with the addition of more short-period CVs, will be 
useful to verify, and perhaps strenghten, the conclusions drawn here.

\section*{Acknowledgments}

We thank an anonymous referee for useful comments and suggestions which 
helped to improve the presentation of our results.
This work is partly based on observations made at Laborat\'orio Nacional
de Astrof\'{\i}sica, Brazil. 
RB acknowledges financial support from CNPq through grant no. 300\,354/96-7. 
BWB acknowledges financial support from CAPES/Brazil.

\bsp

\end{document}